\documentclass[aps,prb,twocolumn,superscriptaddress,showpacs]{revtex4-1}

\usepackage{graphicx} 
\usepackage{wasysym} 

\begin{document}

\title{Dispersive readout of a flux qubit at the single photon level}

\author{J. E. Johnson}
\affiliation{Department of Physics, University of California, Berkeley, California 94720, USA}
\author{E. M. Hoskinson}
\affiliation{Department of Physics, University of California, Berkeley, California 94720, USA}
\affiliation{Quantum Nanoelectronics Laboratory, Department of Physics, University of California, Berkeley, California 94720, USA}
\author{C. Macklin}
\affiliation{Quantum Nanoelectronics Laboratory, Department of Physics, University of California, Berkeley, California 94720, USA}
\author{D. H. Slichter}
\affiliation{Quantum Nanoelectronics Laboratory, Department of Physics, University of California, Berkeley, California 94720, USA}
\author{I. Siddiqi}
\affiliation{Quantum Nanoelectronics Laboratory, Department of Physics, University of California, Berkeley, California 94720, USA}
\author{John Clarke}
\affiliation{Department of Physics, University of California, Berkeley, California 94720, USA}

\date{\today}

\begin{abstract}
A superconducting flux qubit is inductively coupled to a Superconducting QUantum Interference Device (SQUID) magnetometer, capacitively shunted to form a 1.294-GHz resonator. The qubit-state-dependent resonator frequency is weakly probed with a microwave signal and detected with a Microstrip SQUID Amplifier. At a mean resonator occupation $\bar{n} =$ 1.5 photons, the readout visibility is increased by a factor of 4.5 over that using a cryogenic semiconductor amplifier. As $\bar{n}$ is increased from 0.008 to 0.1, no reduction in $T_1$ is observed, potentially enabling continuous monitoring of the qubit state.

\end{abstract}

\pacs{03.67.Lx, 42.50.Pq, 85.25.-j}

\maketitle

Superconducting qubits show promise as scalable building blocks for quantum computing. \cite{Clarke:2008,Neeley:2010uq,DiCarlo:2010kx} A fundamental requirement for attaining this goal is high-fidelity readout of the qubit state. Ideally, this condition is balanced with a readout that minimally perturbs the measured state of the qubit. Dispersive techniques offer the possibility of high repetition rate, quantum nondemolition (QND) readout by avoiding dissipation close to the qubit. \cite{PhysRevA.69.062320,Wallraff_nature,PhysRevLett.101.080502,Manucharyan02102009} A flux qubit coupled to a resonator based on a Superconducting QUantum Interference Device (SQUID) is one architecture that implements this scheme. \cite{PhysRevLett.96.127003,Lupascu:2007} In many instances, readout backaction decreases with lower resonator excitation power. In a low-power analog readout, however, the fidelity is directly limited by the noise performance of the first amplifier in the microwave readout chain, \cite{PhysRevLett.95.060501} traditionally a cryogenic semiconductor high electron mobility transistor (HEMT) amplifier. Low-noise superconducting amplifiers \cite{Bergeal:2010dq,Spietz:2009ly,Castellanos-Beltran:2007ve,Hatridge:2011qf} have been successfully used for applications \cite{Bradley:2003vn,Vijay:2011bh} requiring very high measurement sensitivity. Using a superconducting amplifier as the first stage amplifier decreases the system noise temperature, allowing for shorter integration times and a faster measurement with respect to $T_1$, thus increasing the fidelity.

In this paper, we report the measurement of a superconducting flux qubit using a low noise Microstrip SQUID Amplifier (MSA). \cite{Muck:1998,10.1063/1.3486156} We examine the resultant improvement in qubit readout, and show that it allows practical access to the very weak measurement regime by providing a 4.5-fold increase in the measurement visibility. We further demonstrate that there is no discernible increase in the qubit relaxation rate as the mean resonator occupation $\bar{n}$ is increased from 0.008 to 0.1 photons, opening up the possibility of continuous monitoring of the qubit state with enhanced visibility. Continuous monitoring with high fidelity and minimal readout backaction is necessary for continuous quantum feedback and error correction, \cite{PhysRevA.65.042301} satisfying a requirement for practical quantum information processing.

When biased with an external magnetic flux near the degeneracy point, $\Phi_{\text{ext}} = \Phi_0 / 2$, the potential energy of the flux qubit \cite{vanderWal2000} as a function of the generalized phase coordinate $\phi$ is a double well; $\Phi_0 \equiv h/2e$ is the flux quantum. Classically, the wells are associated with persistent currents in counter-rotating directions, $|\rightturn \rangle$ and $ |\leftturn \rangle$.  Tunneling between the double wells mixes the two circulating states, so that the energy eigenstates of the system are superpositions of $|\rightturn \rangle$ and $ |\leftturn \rangle$. The energy gap $E$ between the ground and excited states is well described by $E = \sqrt{\Delta^2 + \epsilon^2}$, where $\epsilon = 2 I_q \left(\Phi_{\text{ext}} - \Phi_0 / 2 \right)$, $\Delta$ is the minimum energy splitting at the degeneracy point, and $I_q$ is the persistent current magnitude. At the degeneracy point, the qubit eigenstates are equal superpositions of the two circulating current states, differing only by a relative phase. Thus at this bias the two eigenstates are indistinguishable to a measurement which probes the average magnetization of each state.

\begin{figure}
\includegraphics[width=8cm]{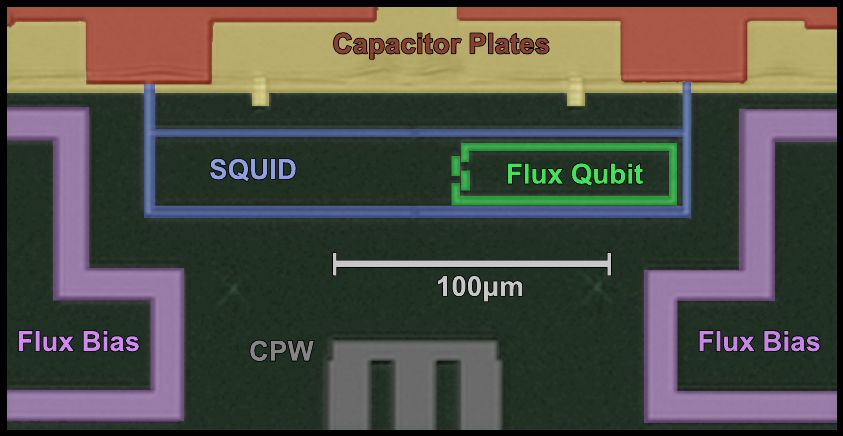}
\caption{(Color online) False color optical micrograph of the qubit and readout resonator.  A three-junction flux qubit is inductively coupled to a dc SQUID shunted by two parallel-plate capacitors in series, forming a tunable nonlinear resonator for qubit readout. The capacitors consist of aluminum films with a niobium underlayer, separated by a $\text{SiN}_x$ dielectric. Symmetric on-chip flux bias lines allow independent tuning of the qubit and SQUID, while a shorted CPW structure allows microwave excitation and fast flux tuning of the qubit.}
\label{fig:falsecolor}
\end{figure}

An optical micrograph of the qubit and associated circuitry is shown in Fig. \ref{fig:falsecolor}.  The device is fabricated on a silicon substrate using double-angle aluminum shadow evaporation. The $22 \times 83~\mu$m$^2$, three-junction flux qubit is coupled via a mutual inductance $M_{QS} = 57$ pH to a $30 \times 200~\mu$m$^2$ dc SQUID, which surrounds the qubit. The qubit parameters are $ \Delta/h = 8.507$ GHz and $I_q = 193$ nA.  The SQUID has a calculated loop inductance $ L = 420$ pH and critical currents of approximately $I_0 = 0.9\ \mu$A per junction. The SQUID is shunted by two on-chip 45-pF capacitors in series; these capacitors share a common niobium base film separated from the aluminum upper films by a 200-nm layer of silicon nitride.

The capacitively shunted SQUID forms a nonlinear resonator with a maximum center frequency of 1.49 GHz, tunable to lower frequencies with an applied flux. The SQUID functions as a nonlinear, flux-dependent inductance. The sensitivity of the resonator frequency to magnetic flux is also the basis for distinguishing the circulating current states of the flux qubit.  The quality factor of the resonator is 10, set by an off-chip surface-mount capacitor that couples it to the readout circuit. Two control lines from an optically isolated current source enable independent flux biasing of the qubit and SQUID. \cite{linzen:2541} Transitions in the qubit are excited via a shorted co-planar waveguide (CPW) which by symmetry does not couple directly to the SQUID.

\begin{figure}
\includegraphics[width=8.6cm]{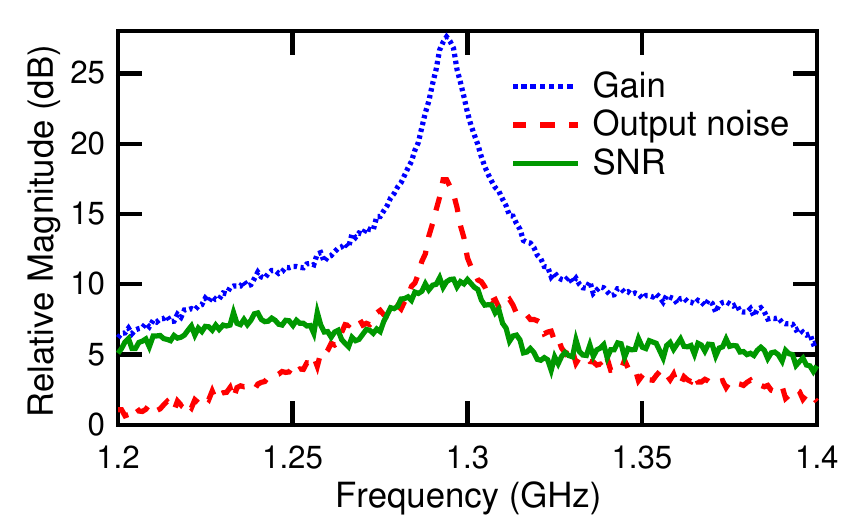}
\caption{(Color online) Improvement in measurement sensitivity due to the MSA. The transmission data are normalized to a configuration with the MSA switched out of the measurement chain. At 1.294 GHz the MSA provides 27-dB gain (blue dots) with close to 10-MHz bandwidth. The increase in the output noise power (red dashes) is 17 dB, yielding an overall improvement in power SNR (increase in ratio of gain to output noise) of 10 dB (solid green line).}
\label{fig:msaperformance}
\end{figure}

The MSA consists of a grounded SQUID washer, with inductance $L_s \approx 450$ pH, electrically isolated by a SiO layer from a strongly coupled, superconducting coil with nine turns. One end of the input coil is left open circuit, defining a $\lambda / 2$ microstrip resonant mode. The input signal is coupled via an off-chip capacitor optimized for critical coupling. The device parameters are comparable to those used in previous MSAs with two notable improvements intended to enhance performance at higher frequencies. \cite{kinion:172503,kinion:202503} First, a large niobium pad that connects to the washer and is wirebonded directly to the ground of the printed circuit board decreases the inductance of the SQUID washer to ground. Second, the input coil linewidth is narrowed to 1 $\mu$m using e-beam lithography. This reduces the coil-washer capacitance while only slightly increasing the inductance, which is dominated by the inductive loading of the SQUID. \cite{Ketchen:736} The net result is an increase in the microstrip resonance frequency. The SQUID is current biased into the voltage state and flux biased near $\Phi_0 / 4$ to yield an optimal flux-to-voltage transfer function $V_{\Phi} \equiv \partial V / \partial \Phi$. Flux coupling to the SQUID is strongly enhanced on resonance, leading to the optimized amplification characteristics shown in  Fig. \ref{fig:msaperformance}. Maximum performance is achieved on resonance at 1.294~GHz, where the amplifier exhibits a stable power gain of 27 dB with a bandwidth of almost 10 MHz. This gain is accompanied by a rise in the system noise of 17 dB, so that the MSA provides a signal-to-noise ratio (SNR) increase of 10 dB.  The amplifier is linear until the input reaches a level of -124 dBm, corresponding to $\bar{n} \approx$ 0.6 in the readout resonator, where the output signal is compressed by 1 dB. Though the compression point of this MSA is lower than usual, any SQUID-based amplifier is intrinsically limited to flux signals of less than $\Phi_0/2$.

\begin{figure}
\includegraphics[width=8cm]{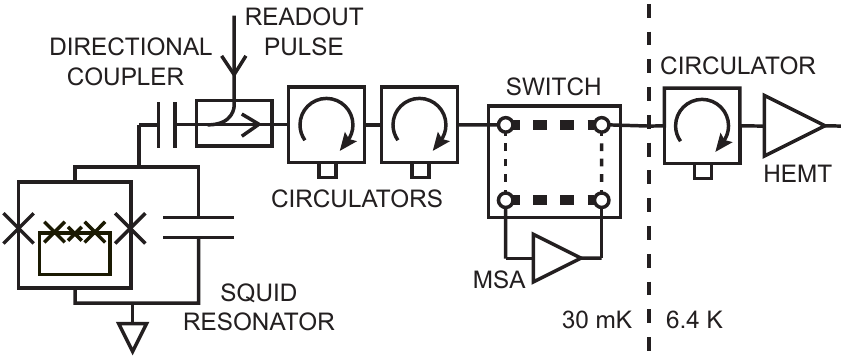}
\caption{Schematic of cryogenic microwave measurement chain. A switch anchored to the mixing chamber stage allows the biased and operating MSA to be switched into/out of (thin/heavy dashed lines only) the readout circuit, allowing for a direct comparison of both configurations.}
\label{fig:circuitschematic}
\end{figure}

To measure the qubit state with the MSA we use standard homodyne detection. A schematic of the cryogenic part of the readout signal chain is shown in Fig. \ref{fig:circuitschematic}. The qubit is initially biased at the degeneracy point, where the $\pi$-pulse fidelity is highest and most insensitive to flux drifts. The relaxation ($T_1$) and dephasing ($T_2^*$) times at this point are 320 ns and 250 ns, respectively. We apply either a $\pi$ pulse to prepare the qubit in the excited state or no pulse to leave it in the ground state. At the conclusion of the qubit control pulse, a fast flux shift is applied to move the ground and excited states adiabatically into differing superpositions of circulating current states, producing the magnetization signal between the qubit states. The qubit pulses and fast flux shift are added together using a directional coupler at base temperature, and are connected to the on-chip, shorted CPW. The readout pulse, shaped at room temperature, is coupled to the resonator via a separate directional coupler. This signal is reflected by the SQUID resonator, thus encoding the qubit circulating current state in the reflected phase. The resonator frequency is set to $f_r =$ 1.294 GHz, matching the operating frequency of the MSA, by flux biasing the SQUID at $\Phi = 0.42 \Phi_0$, at which point $d f_r / d \Phi$ is high. The reflected signal is amplified by the MSA and HEMT as it passes up the output chain, which includes a network of circulators and low loss niobium coaxial lines. After further amplification at room temperature the readout signal is mixed down with the carrier signal and the in-phase and quadrature signals are digitized.

We first measure the fidelity at a microwave power sufficiently high to explore the nonlinearity of the resonator. In this regime ($\bar{n} \approx $ 500), we exploit bifurcation to achieve a latching readout, \cite{PhysRevB.73.054510,Mallet:2009oq} resulting in a fidelity [$1-P_{0|1}-P_{1|0}$, where $P_{0|1} \ (P_{1|0})$ is the fraction of erroneous excited-(ground-) state counts] of 67\%. Our focus in this paper, however, is the low-power, analog region where the absolute fidelity is lower but the resonator response is very nearly linear.

The resonator phase difference between $|\rightturn \rangle$ and $ |\leftturn \rangle$ is $19^{\circ}$\textemdash the maximum signal we expect in the absence of all losses. In our experiment, we measure the phase difference at the instant readout begins by monitoring the ensemble average phase difference between typically $10^5$ ground- and excited-state preparations as it exponentially decays to zero during readout. Fitting these data and extrapolating back to time $t=0$ (termination of $\pi$ pulse and initiation of readout) reveals a maximum phase contrast of $8.5^{\circ}$, produced by a flux shift of  $\sim$4 m$\Phi_0$. During the shift, there is a loss of excited-state population that we postulate occurs as the qubit is swept through strongly coupled environmental decay modes. Below 4 m$\Phi_0$, the gain in fidelity due to the increasing flux shift outweighs this population loss, while the opposite is true above 4 m$\Phi_0$. This implies that, at the optimal flux shift, the qubit eigenstates at readout contain non-negligible superpositions of the two circulating current states, preventing ideal mapping of the qubit eigenstate onto the circulating current readout basis. Ultimately, these effects before readout limit the maximum possible fidelity to 45\%, the ratio of the measured ($8.5^{\circ}$) to ideal ($19^{\circ}$) phase differences.

Processes during readout also contribute to fidelity loss. First, the limited bandwidth of the MSA leads to a 50-ns delay\textemdash during which $T_1$ decay events occur\textemdash between the start of readout and the time at which the average signal between ground- and excited-state readouts peaks. After accounting for the circulating current admixture, population loss during the fast flux shift, and $T_1$ decay, the SNR contribution is then inferred as the remaining fidelity loss. Table \ref{tab:fidlosstab} tabulates the relative contributions of these effects.

\begin{table}
\begin{ruledtabular}
\caption{Accounting of the sources of fidelity loss at $\bar{n} = $ 1.5 photons with MSA readout. The measured fidelity is 27.7\%. The right column accounts for the cumulative fidelity loss at and above a given row.}
\label{tab:fidlosstab}
\begin{tabular}{ c  c  c  }
Source of nonideality & Fidelity loss & Max measurable \\
& & fidelity \\
\hline
Flux shift \& decay & 55\% & 45\% \\
$T_1$ decay & 15\% & 38\% \\
SNR & 28 \% & 27.7\% \\
\end{tabular}
\end{ruledtabular}
\end{table}

\begin{figure}
\includegraphics[width=8.6cm]{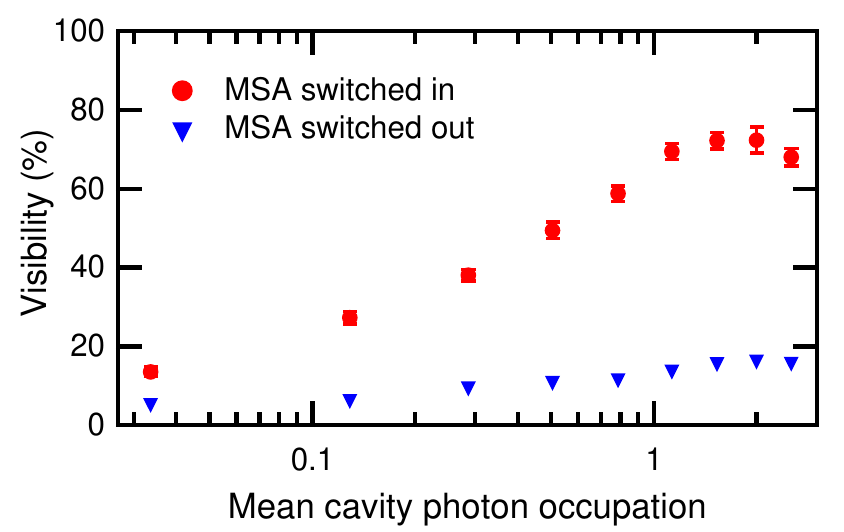}
\caption{(Color online) Visibility vs mean cavity photon occupation for the MSA switched into and out of the circuit. A visibility of 100\% corresponds to the calculated maximum achievable fidelity limited by all factors except SNR. The error bars on the blue triangles are smaller than the markers.}
\label{fig:visibility}
\end{figure}

To separate the performance of the readout from nonidealities of the qubit, we define the visibility as the raw fidelity normalized to the maximum fidelity determined by all sources of fidelity loss except finite SNR, as shown in Table \ref{tab:fidlosstab}. Thus, the visibility is not a direct function of the qubit operating point. The visibility as a function of the number of readout photons in the cavity with the MSA switched in and switched out of the circuit is shown in Fig. \ref{fig:visibility}. The raw fidelity is calculated by optimally postprocessing the readout signal, which entails weighting each readout trace with an exponentially decaying filter function, integrating the result, \cite{exponential_filter} and histogramming the response of $10^5$ ground- and $10^5$ excited-state preparations. For low-amplitude excitation, as expected, the visibility is very low, and the qubit states are virtually indistinguishable. As the number of photons is increased, the SNR and thus the visibility increase. The addition of the MSA substantially increases the visibility.  At 1.5 photons, the visibility with the MSA reaches $72.3 \pm 2.1 \%$ ($27.7 \pm 0.7 \%$ fidelity), a factor of $\sim$4.5 greater than without the MSA. At higher photon numbers, the visibility decreases as the nonlinearity in the SQUID resonator causes the resonance to shift to lower frequencies. Overall, the measured visibility is restricted by the relatively small $19^{\circ}$ phase shift in the resonator between the two circulating states, a consequence of the low quality factor of the resonator.

We independently confirm that the measured visibility and corresponding fidelity loss due to SNR at $\bar{n} =$ 1.5 are consistent with the measured system noise, bandwidth, and resonator phase difference between $|\rightturn \rangle$ and $ |\leftturn \rangle$. We simulate the resonator response of $10^6$ excited- and ground-state readouts using these measured properties. The excited-state records are allowed to decay stochastically on an exponential time scale given by the measured $T_1$. The visibility is calculated after these readouts are filtered, integrated, and histogrammed in an identical manner to the measured data. This calculation predicts a visibility of 76\% at $\bar{n} =$ 1.5, in very reasonable agreement with the measured value of 72.3\%.

\begin{figure}
\includegraphics[width=8.6cm]{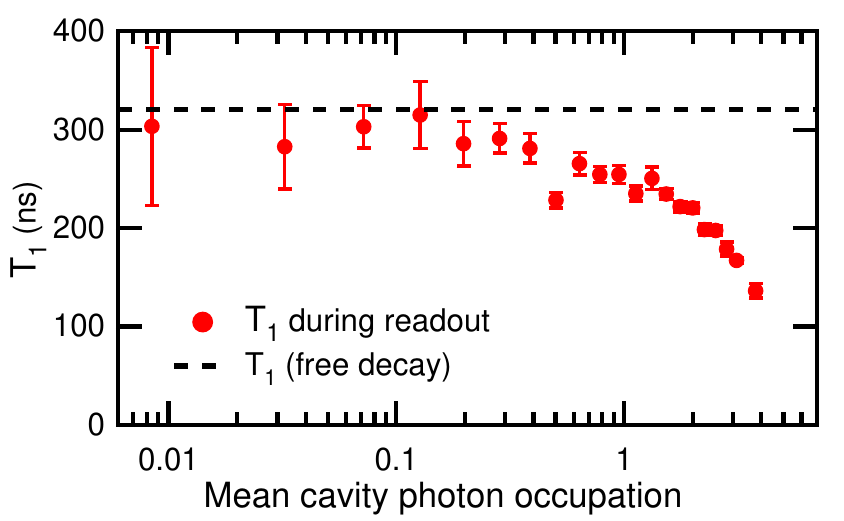}
\caption{(Color online) Relaxation time constant ($T_1$) during readout vs mean cavity photon occupation.}
\label{fig:T1duringread}
\end{figure}

In general, the visibility of dispersive readout is limited by the noise floor of the following amplifier, so that using higher photon numbers to readout the qubit state should increase the visibility. \cite{PhysRevLett.95.060501} Any backaction associated with the measurement also increases with $\bar{n}$, however, so there should be an optimum measurement strength that balances visibility against backaction. To study the effect of measurement strength on qubit-state evolution, we measured the relaxation time $T_1$ during continuous readout monitoring versus $\bar{n}$ (Fig. \ref{fig:T1duringread}).  We extract $T_1$ during readout by fitting the average phase difference between ground- and excited-state preparations versus time.  For 0.008 $< \bar{n} <$ 0.1 during readout, $T_1$ approaches 320 ns, the value measured by applying a $\pi$ pulse and waiting a variable delay time before readout (free decay). As cavity occupation increases, $T_1$ decreases. We speculate that this reduction in $T_1$ with increasing readout power arises from sweeping the qubit through increasing numbers of environmental decay modes. As the readout power is increased above $\bar{n}$ = 4, the frequency of the nonlinear SQUID resonator decreases and its response sharpens. Consequently, the fraction of incident power entering the resonator decreases, and our photon number calibration fails as $\bar{n}$ no longer increases linearly with incident power.

In conclusion, we have successfully integrated a MSA into a qubit measurement scheme with a substantial corresponding increase in readout visibility.  This improvement is particularly beneficial at low readout powers where readout-induced reduction of $T_1$ is minimal and the MSA is operated within its optimum dynamic range. This improved visibility allows access to the very weak continuous quantum measurement regime \cite{weak_measurement} while preserving sufficient SNR to resolve the qubit state more efficiently with minimal readout backaction.  Coupled with a qubit readout architecture optimized for high readout phase contrast (i.e., by increasing the resonator quality factor by engineering for a smaller coupling capacitance), real-time monitoring of the qubit state with very high visibility should be achievable. The MSA offers the advantage of requiring only static current and flux biases, \cite{Muck:1998,10.1063/1.3486156,Spietz:2009ly} greatly simplifying the microwave infrastructure required to operate the amplifier. Furthermore, although circulators and directional couplers were used in this experiment, in principle the forward directionality of the MSA eliminates the need for such nonreciprocal components when paired with a transmission-based readout resonator. This opens up the possibility of on-chip lithographic integration, which, needless to say, would require careful engineering of the resonator output and MSA input impedances. The direct coupling would also allow for an investigation of the MSA backaction noise spectrum\textemdash notably at the Josephson frequency\textemdash and its effect on qubit coherence.

\begin{acknowledgments}
We thank R. Vijay, D. Kinion, O. Naaman, D. Murray, and J. Jeet for useful discussions and contributions to this project. D.H.S. acknowledges support from a Hertz Foundation Fellowship endowed by Big George Ventures. This work was funded in part by the U.S. Government and by BBN Technologies.
\end{acknowledgments}



%

\end{document}